\documentclass[manuscript,screen]{acmart}
\AtBeginDocument{%
  }

\setcopyright{rightsretained}
\copyrightyear{2026}
\acmYear{2026}
\acmConference[CHI '26]{CHI 2026 Workshop on
Human-Agent Collaboration}{April 13–17, 2026}{Barcelona, Spain}
\acmBooktitle{CHI 2026 Workshop on Human-Agent Collaboration, April 13--17, 2026, Barcelona, Spain}
\acmISBN{978-1-4503-XXXX-X/2018/06}

\usepackage{enumitem}

\usepackage{xspace}

\newcommand{\paratitle}[1]{\vspace{1.0ex}\noindent\textbf{#1}}
\newcommand{\ie}{\textit{i.e.,}~}
\newcommand{\eg}{\textit{e.g.,}~}

\begin{document}

\title{From Control to Foresight: Simulation as a New Paradigm for Human–Agent Collaboration}


\author{Gaole He}
\email{hegaole@nus.edu.sg}
\affiliation{%
 \institution{National University of Singapore}
 \city{Singapore}
 \country{Singapore}
}

\author{Brian Y. Lim}
\email{brianlim@nus.edu.sg}
\affiliation{%
 \institution{National University of Singapore}
 \city{Singapore}
 \country{Singapore}
}

\renewcommand{\shortauthors}{He et al.}

\begin{abstract}
  Large Language Models (LLMs) are increasingly used to power autonomous agents for complex, multi-step tasks. 
  However, human-agent interaction remains pointwise and reactive: users approve or correct individual actions to mitigate immediate risks, without visibility into subsequent consequences. 
  This forces users to mentally simulate long-term effects, a cognitively demanding and often inaccurate process. 
  Users have control over individual steps but lack the foresight to make informed decisions. 
  We argue that effective collaboration requires foresight, not just control. 
  We propose simulation-in-the-loop, an interaction paradigm that enables users and agents to explore simulated future trajectories before committing to decisions. Simulation transforms intervention from reactive guesswork into informed exploration, while helping users discover latent constraints and preferences along the way. This perspective paper characterizes the limitations of current paradigms, introduces a conceptual framework for simulation-based collaboration, and illustrates its potential through concrete human-agent collaboration scenarios.
\end{abstract}





\maketitle

\section{Introduction}
Equipped with toolkits such as browser search and machine access, large language models (LLMs) have shown promising potential to interact with the world and assist with more complex, multi-step tasks---ranging from travel planning to code generation~\cite{wang2024survey,xi2025rise}. 
To ensure reliable and accountable outcomes, LLM agents are often supervised by humans~\cite{Feng-cocoa-2025,he2025plan}: the agent proposes a sequence of actions, and the human is asked to approve or correct decisions at key junctures. 
However, this form of collaboration rests on an implicit assumption: that humans can act as oracles~\cite{Stephan-AAAI-2023,Katherine-AIES-2023}, making sound decisions with minimal context. 
In practice, when humans are inserted into multi-step workflows, they are asked to make critical decisions with limited information---without visibility into how their approval/action might shape subsequent outcomes~\cite{Maurice-arxiv-2025}.
Without visibility into downstream consequences, human intervention becomes short-sighted rather than informed collaboration~\cite{kolt2025governing}. 
In these cases, humans are forced to rely on intuition or mental simulation, both of which are prone to error---especially as task complexity grows.

This limitation is especially pronounced in long-horizon tasks, where early decisions cascade into future outcomes in ways that are difficult to anticipate~\cite{he2025fine}. 
For example, a seemingly plausible choice (\eg booking a tight connection flight) can propagate through the planning horizon, amplifying or constraining downstream possibilities. 
Yet current human-agent interactions do not proactively support reasoning about these ripple effects. 
Without such support, users are left to mentally simulate alternative futures, a process that is cognitively demanding and notoriously unreliable, especially when tasks involve long-range dependencies and stochastic outcomes. For instance, a delayed flight can cascade into missed connections. 
Worse still, this narrow view forecloses serendipity~\cite{SalmaHP25}: when humans see only the immediate next step and react to it, they miss the opportunity to discover unexpected but valuable alternatives that lie off the agent's proposed path. 
This results in a substantial asymmetry: while LLM agents can explore possible actions (\eg tree-based search over the action space~\cite{koh2025tree,Chen-IJCAI-2024}) and their subsequent impacts, the human collaborator is given access to only a single path through that tree---the trajectory proposed by the agent.

We argue that addressing this asymmetry requires more than just giving humans control over actions---it requires giving them foresight into the consequences of those actions. 
Control without foresight is like driving at night with no headlights: you can turn the wheel, but you cannot see what lies ahead. 
To this end, we propose simulation-in-the-loop collaboration, an interaction paradigm that allows humans and agents to preview counterfactual future trajectories before committing to decisions. 
By generating and visualizing possible outcomes across multiple paths, simulation transforms human intervention from reactive guesswork into proactive exploration. 
It also creates space for serendipity: as users explore simulated futures, they may discover valuable alternatives or latent constraints that were not apparent from the agent's initial proposal. 
In this paper, we (1) articulate the limitations of existing pointwise interaction paradigms, (2) introduce a conceptual framework and design space for simulation-based collaboration, and (3) illustrate the approach through concrete scenarios with LLM agents and other planning tasks.


\section{Simulation-in-the-loop Collaboration}
We introduce simulation-in-the-loop collaboration, an interaction paradigm in which humans and agents jointly explore simulated future trajectories before committing to real-world actions. 

\subsection{Concept and Definition}
First, we introduce four core concepts that ground our framework:

\begin{itemize}[noitemsep, topsep=0pt, partopsep=0pt, parsep=0pt]
    \item \textbf{Agentic Workflow}: The unit of analysis is a multi-step task performed by an LLM agent under human oversight (\eg travel planning). The workflow proceeds through a sequence of actions, during which the agent may request human input. For simplicity, we can view it as a step-by-step planning-execution process~\cite{he2025plan,Feng-cocoa-2025}: at each step, the agent predicts an action, and the human may approve, modify, or override it before execution.
    \item \textbf{Action Space}: At each step, the agent considers multiple possible actions (\eg search flights, search relevant news, and search social media), each leading to different downstream trajectories. LLM agents inherently explore such trees during planning---through beam search~\cite{SutskeverVL14}, Monte Carlo tree search~\cite{chen2025human,Chen-IJCAI-2024}, or implicit generation of alternatives---but this exploration typically remains internal and invisible to the human.
    \item \textbf{Simulation}: We define simulation as the agent's ability to externalize this internal exploration: before committing to a decision, the agent generates and presents multiple future trajectories for human preview. Simulation is not planning (which seeks an optimal path) but rather exploration for sensemaking---making the tree of possibilities visible and navigable.
    \item \textbf{Simulated Impact}: Each simulated trajectory is annotated with key outcomes---risks, opportunities, trade-offs, uncertainties---that help humans compare alternatives. Simulated impact translates abstract futures into concrete, decision-relevant outcomes, enabling foresight rather than guesswork.
\end{itemize}

\begin{figure}[htbp]
    \centering
    \includegraphics[width=0.9\textwidth]{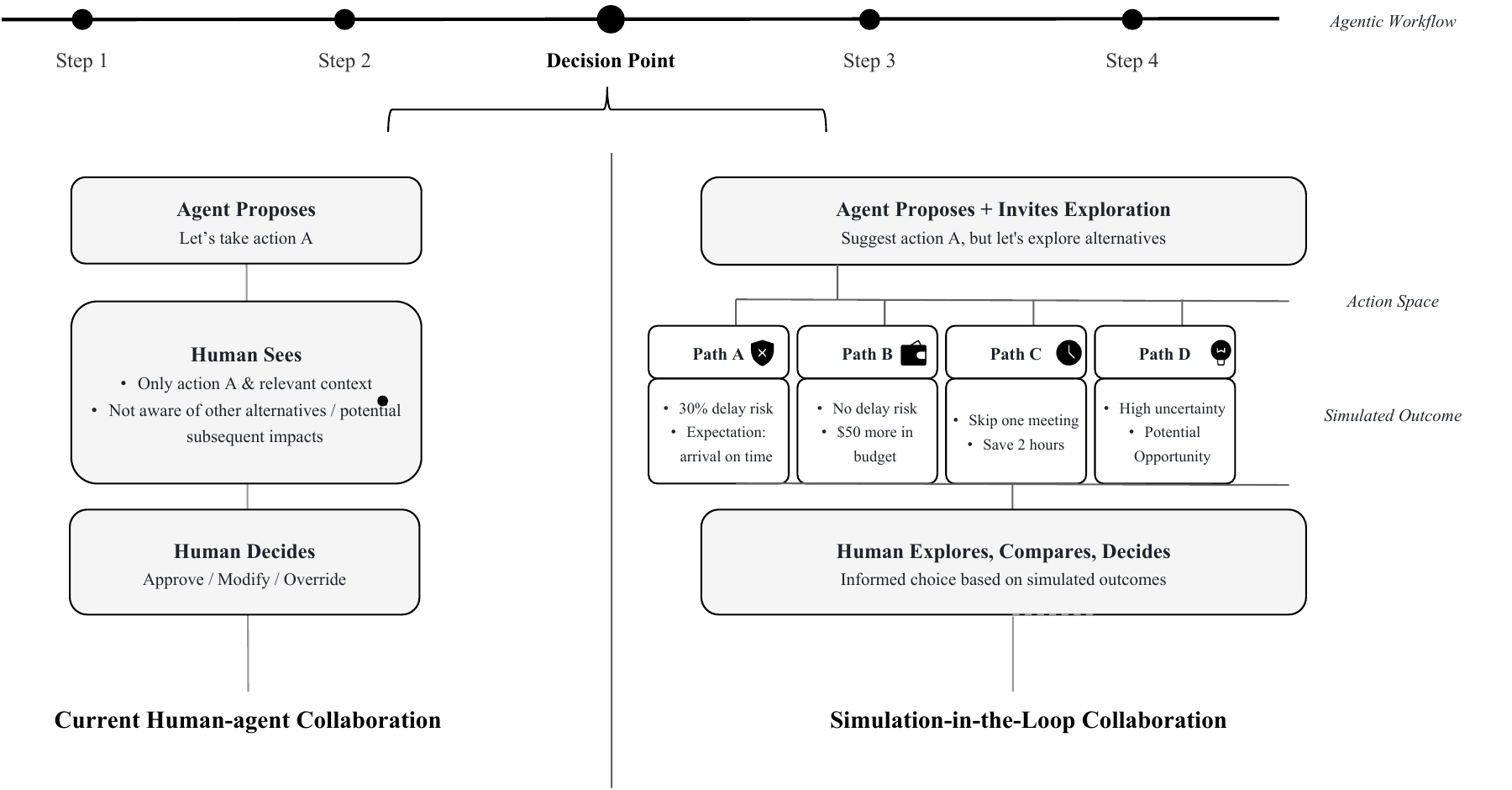}
    \caption{Comparison of interaction paradigms at a decision point.}
    \label{fig:concept}
    \Description{(Left) Traditional pointwise collaboration: the human sees only the agent's proposed action and must decide without downstream visibility. (Right) Simulation-in-the-loop: the agent externalizes multiple possible futures with simulated impacts, enabling informed comparison and serendipitous discovery.}
\end{figure}

\subsection{Illustrative Scenario}
We ground our framework in one concrete scenario---a multi-city trip planning. As visualized in Figure~\ref{fig:concept}, the simulation acts as an intermediate layer between agent proposals and human commitment. 
At a decision point, the agent must choose whether to book a tight connection between flights. 

In the current human-agent collaboration mode, the agent would simply propose its preferred option \ie Path A: a direct flight with a 1-hour layover---and ask for feedback. 
The user, seeing only that this option is available and cheaper than alternatives, might approve without realizing the downstream risk or that lower-risk alternatives even exist.

With simulation-in-the-loop, the agent externalizes multiple alternatives (Paths A-D in Figure~\ref{fig:concept}). 
Path A (the agent's proposal) is annotated with a simulated impact: $30\%$ delay risk due to short connection time. 
Path B, a later flight with a longer layover, shows no delay risk but adds $\$50$ cost. 
By comparing these simulated outcomes, the human discovers a risk they hadn't considered and can make an informed trade-off between time and reliability. 
The simulation also reveals Path D: a flight into a different airport, which opens up new options the user hadn't considered, illustrating how simulation enables serendipitous discovery. 
By externalizing possible futures, the human’s role shifts from reactive supervision to proactive planning and negotiation with the agent.

\subsection{Design Space for Simulation-Based Interaction}

Designing effective simulations involves navigating trade-offs between three key dimensions. These dimensions are not merely technical parameters. They shape how humans reason, trust, and collaborate with agents---making simulation a design choice.

\paratitle{Lookahead Depth}. How far into the future should simulations project? Deeper lookahead provides greater foresight but risks information overload and compounding uncertainty. Shallower previews are more reliable but may miss critical downstream effects.

\paratitle{Exploration Breadth}. How many alternative futures should be shown? A single trajectory minimizes cognitive load but risks tunnel vision. Multiple branches enable comparison and serendipity, but may overwhelm. Beyond quantity, systems must also convey outcome diversity---ensuring displayed futures represent meaningfully different possibilities.


\paratitle{Granularity}. How detailed should simulations be? Fine-grained simulations (\eg code execution) provide rich information but incur latency. Coarse-grained approximations (\eg LLM sketches) are faster but risk omitting critical details or hallucinating. The balance depends on task criticality.






\section{Challenges and Opportunities}

Implementing simulation-in-the-loop collaboration introduces several challenges, yet also opens new directions for human-agent collaboration.

\subsection{Challenges}

\paratitle{Simulation Reliability}. Simulation requires a model or environment that can generate plausible future trajectories. While this is feasible for tasks with well-defined environments (\eg games and code execution), it becomes technically challenging in open-ended domains where world dynamics are less structured. 
Relying on LLMs to simulate their own futures---asking the agent to predict ``what if''---offers a tempting but uncertain alternative. 
LLM-generated simulations may hallucinate, omit critical dependencies, or produce overly optimistic trajectories~\cite{WangTXYCCJ24,gao2024large,grattafiori2025word}. 
This highlights a need for more reliable world models~\cite{pan-abs-2511-09057} that can support simulation across diverse, open-ended tasks.

\paratitle{What to Simulate}. Not all possible futures are valuable for users' attention. Simulations can generate an abundance of trajectories, but presenting trivial or near-identical options. Thus, it is important to identify which outcomes are non-trivial and decision-relevant: surfacing paths that reveal genuine trade-offs, hidden risks, or unexpected opportunities, while filtering out those that offer no new insight.

\paratitle{Cognitive Load}. Even with careful filtering, comparing multiple futures imposes cognitive demands. The interface is supposed to help users navigate across trajectories, understand trade-offs, and track their exploration—without becoming a source of confusion itself. If users cannot integrate simulated outcomes into their decisions, simulation adds noise rather than insight.






\subsection{Oppotunities}

\paratitle{From Reactive to Proactive Collaboration}. Current proactive agents act autonomously unless interrupted. Simulation-in-the-loop offers a middle ground: agents proactively show possible futures, inviting human input before acting. This shifts collaboration from ``human as supervisor'' to ``human as explorer''.

\paratitle{Enabling Backtracking by Anticipation}. Recent work on backtracking agents~\cite{wu-etal-2025-backtrackagent} focuses on error detection and recovery in LLM agents. Simulation flips this backward-looking repair into forward-looking prevention---helping humans and agents avoid dead ends before committing.


\paratitle{Discovering Latent Constraints and Needs}. As users explore simulated futures, they encounter constraints embedded in the task—dependencies, resource limits, timing conflicts---that were not visible from the initial proposal. At the same time, they may discover gaps between their expectations and what is achievable, revealing unstated preferences or new goals. This turns collaboration into joint discovery: requirements emerge dynamically through exploration, and the agent's final outcome improves precisely because these latent factors are surfaced before commitment, not after.







\bibliographystyle{ACM-Reference-Format}
\bibliography{simulation}


\end{document}